\documentclass[10pt]{article}
\newcommand{\ber}{\begin{eqnarray}}
\newcommand{\eer}{\end{eqnarray}}
\newcommand{\bea}{\begin{equation}}
\newcommand{\eea}{\end{equation}}
\newcommand{\f}[1]{\frac{#1}{2}}
\def\be{\begin{eqnarray}}
\def\ee{\end{eqnarray}}
\def\bd{\begin{displaymath}}
\def\ed{\end{displaymath}}

\def\ga{\gamma}
\def\al{\alpha}

\def\ra{\rightarrow}
\def\etal{{\em et al} }
\def\PR{{\em Phys. Rev. }}
\def\NP{{\em Nucl. Phys. }}
\def\jpg{{\em J. Phys. G: Nucl. Part. Phys. }}
\def\PL{{\em Phys. Lett. }}
\def\ZP{{\em Zeit. Phys. }}
\def\APNY{{\em Ann. Phys. ,N.Y. }}

\begin{document}
\title{Relativistic mean 
field study of the newly discovered $\alpha$-decay chain of $^{287}$115 }

\author{Sankha Das\\
Department of Physics, MCKV Institute of Engineering,\\
243 G.T. Road (N), Liluah, Howrah 711 204, India\\
and\\
 G Gangopadhyay \\
 Department of Physics,
 University of Calcutta,\\
92, Acharya Prafulla Chandra Road, Kolkata-700 009, India}
\date{\today}
\maketitle

\begin{abstract}
The single particle structure of the low lying levels in the nuclei 
belonging to the newly observed $\alpha$-decay chain starting from $^{287}$115
has been studied. Relativistic mean field calculation with two forces
NL1 and NL3 has been  performed in the blocked BCS approximation. Q-values
for possible decay have been calculated and compared with experiment. 
Lifetime estimates obtained from phenomenological relations have also
been compared.
\end{abstract}

\section{Introduction}

One of the most exciting fields in nuclear physics in recent times is the 
production
and study of superheavy nuclei. It has been possible to synthesize nuclei 
up to $Z$=116 in the laboratory. The most recent additions are the nuclei 
$^{287,288}$115\cite{115}. The odd nucleus $^{287}$115 has been observed to 
decay through $\al$-emissions  and most probably ending in the nucleus 
$^{267}$105. 
Superheavy nuclei present an exciting challenge to the existing nuclear 
theories as their very existence is a quantum effect, the liquid drop barrier 
being either absent or negligibly small. Theory plays a very important role 
also in the
identification of nuclei in the superheavy region.

In any study of odd mass nuclei in the superheavy region the most
important quantities are the binding energy and the $Q_\alpha$ value.
In many cases, the binding
energy is known from the $Q_\alpha$ values of the decays of the chain when
the mass of any member of the chain is known previously, assuming the decay
to be between the ground states of the parent and the daughter nuclei.
However, decay from ground state to ground state may be hindered or
forbidden due to spin-parity selection rules. Thus
it is important to study the one quasiparticle states near the ground
state for possible decay paths. If the observed decay involves isomeric
excited states, the $Q_\alpha$ value may differ considerably from that 
expected in ground state to ground state transition. 
Frequently the chain does not end in any previously known nucleus making the
identification of the parent very difficult. The Q-values for $\al$-decay
and the lifetime estimates obtained from them using 
phenomenological relations are used for the identification. 
As already mentioned, the hindrance factors of $\al$-decay, and
hence lifetime values, depend 
crucially on the quantum numbers of the parent and daughter states involved.
Theoretical ideas about the single particle structure are therefore crucial
even for identification.

Relativistic mean field (RMF) calculations are known to give a good 
description of the structure of nuclei throughout the periodic table. 
In an odd nucleus, there is an additional complication as the last odd 
nucleon breaks the time reversal symmetry of the mean field. Hence, the
space-like components of the meson fields, which are absent in even-even 
nuclei,  contribute to the different nuclear properties in odd nuclei. 
However, it is well known that these contributions are very small in the 
case of bulk properties like binding energy or deformation and are important 
only in features
like magnetic moment or moment of inertia which depend on only a few
nucleons\cite{H,A}.  We have neglected these contributions as we are primarily
interested in only the binding energy. We have restored
time reversal symmetry by using the blocking method in pairing
\cite{Schuck}.

Proton-odd neutron-even superheavy nuclei have not been extensively studied in 
the mean field formalism. Ren \etal\cite{ren} have  systematically 
studied the binding energy and deformation of odd $Z$ superheavy nuclei 
using the relativistic forces TMA\cite{TMA} and NLZ2\cite{NLZ2}.
Rashdan \cite{rash} have studied the nuclei $^{271}$111 using the force NLRA1.
These calculations do not take the level occupied by  last odd nucleon into
account. Geng \etal\cite{geng} have studied the the nuclei of the $\al$-decay chain
starting from $^{287}$115 using the TMA force and included the effect of 
the blocking by the odd proton. They have used both constant pairing and 
density dependent delta-function interaction.

In this work we have studied the structure of the nuclei belonging 
to the decay chain of $^{287}$115.
First we investigate the heaviest proton-odd nuclei whose ground state 
spin parity values are known to check the ability of the theory to reproduce
the experimental structure in superheavy region. Next we study the nuclei 
occurring in the $\al$-decay chain of $^{287}$115 to find out the energy 
and spin-parity of the one quasiparticle levels near the ground state 
and hence the possible $\al$-transitions. We have used the relativistic 
forces NL1\cite{NL1} and NL3\cite{NL3}.

\section{Results}

The RMF theory is well known and will not be discussed here. 
Readers may
find excellent descriptions of the theoretical method using the
basis expansion method in refs \cite{gamb,ring}. We have described the blocking
procedure followed in RMF in an earlier calculation \cite{parna}.
The levels are denoted by the Nilsson quantum numbers. The number of oscillator
shells used in both the fermion and the boson sectors is 22. Pairing has been 
included in the constant gap approximation with the gap being given by
$\Delta_{n,p}=11.2/\sqrt{A}$.

Table \ref{be} presents the calculated binding energy and quadrupole 
deformation of the ground state and the excitation energy and spin-parity 
of the lowest one quasiparticle states obtained in three Md nuclei using the 
forces NL1 and NL3. These are the heaviest odd-proton
nuclei whose ground state spin-parity values are experimentally known. 
 The deformation parameters of
the excited states are very close to that of the ground state and have not
been shown. One can see that either the ground state spin is either correctly
predicted or the state corresponding to the experimental ground state lies
within approximately 0.5 MeV in all cases. This is comparable to the results 
we have obtained earlier in a lower mass region\cite{parna}.

Next we study the odd-proton nuclei belonging to the  chain $^{287}$115.
These are the heaviest odd proton nuclei that have been experimentally 
observed. Nothing is known about the spin-parity of the  low-lying levels 
in these nuclei. In table \ref{be2} the binding energy values calculated using 
the forces NL1 and NL3 for the lowest energy level i.e. ground state in 
different nuclei are given. In table \ref{lev} we present  the results of 
our calculation for the lowest lying one quasiparticle levels.
There are both normal negative parity states as well as positive parity
states arising out of the intruder orbital in the low excitation
energy region and a pattern of low-lying 
isomeric states are expected. For example, in $^{287}$115, the $\f{11}^+$ state
is expected to be an isomeric state in both the calculations. 
Similarly, in $^{267}$Db, the lowest lying negative-parity state is likely 
to be isomeric. Thus one may have two groups of $\al$-particles
connecting the different parity levels among themselves which are nearly
equal in energy and hence comparable lifetime values. An isomeric state may
also decay to the ground state by $\ga$-emission. In most of the nuclei studied 
the ground state spin predicted by the two forces agree with each other 
lending confidence to the ability of RMF to predict the spin-parity of
the correct ground state in this mass region.

The present results differ in one  important detail from that of 
Geng \etal\cite{geng}. In their work the calculated quadrupole 
deformation values decrease smoothly from 0.22 in $^{267}$Db to 0.17 in
$^{283}$113. However, at $^{287}$115, this suddenly jumps up to 0.50.
Any abrupt change in deformation value should translate in an 
increase in $\al$-lifetime. However, as discussed later, experimental lifetime 
estimates do not show any large increase in lifetime for decay from $^{287}$115.
The calculated quadrupole deformation parameter for both the forces in our 
calculation, on the other hand, vary smoothly from 0.28 to 0.17 approximately  
with increase in mass number. This decrease may possibly be ascribed to the 
shell closure at 
neutron magic number $N$ = 172 in agreement with other
relativistic calculations \cite{sh1,sh2} which predict the neutron and proton 
shell closures at 172 and 120, respectively. It is also known that $Z$ = 114 shows 
a small gap which is manifested here in the larger excitation energy values of 
the quasiparticle states in $^{287}$115, particularly for the calculation 
using the NL3 force.

A particular level may decay either via $\al$-decay or 
$\ga$-emission. Ground state, and frequently,  isomeric states will undergo
 $\al$-decay. In table \ref{Q} the theoretical Q-values for $\al$-decay 
are given for different levels. 
The calculated Q-values have been taken assuming that $\al$-decay takes place
between states having the same parity and nearly equal spin. The hindrance 
factors for these transitions are expected to be close to one. In $^{287}$115, 
the $\al$-decay may even start from the isomeric positive parity state. 
Q-value for any other transition may be easily calculated from tables \ref{be2}
and \ref{lev}. The experimental Q-values appear to agree very well with theory,
particularly with the transitions involving negative parity states calculated 
using the force NL3.

In table \ref{T}, we compare the experimentally obtained lifetime 
 with the phenomenological estimate derived from the theoretical
$Q_\al$-value obtained using the NL3 force for the transitions involving
negative parity states given in the first row of table \ref{Q}. All these 
transitions are between states having the same spin-parity and should have 
low hindrance factor. The Viola-Seaborg formula\cite{VS}
\be {\rm log}_{10}(T_\al)=(aZ+b)Q_\al^{-1/2}+cZ+d\ee
where $T_\al$ are given in seconds and $Q_\al$ in MeV, has been used. The 
values of the 
parameters $a,b,c$ and $d$ are taken from Ref. \cite{M} where they were 
obtained by fitting the experimental data of middle and heavy nuclei. 
The values are $a$=1.66175, $b$= -8.5166, $c$= -0.20228 and $d$= -33.9069.
Except for the transition from $^{283}$113, all the other $T_\alpha$ values
agree with experimentally observed values indicating that the hindrance 
factor is close 
to one, i.e. the transitions are between states having nearly equal spin and
the same parity. As mentioned earlier, the lifetime estimate of $\al$-decay 
from $^{287}$115 does not indicate any large difference in deformation values 
between the parent and the daughter states as predicted by Geng \etal 
\cite{geng}.

\section{Summary}

The structure of superheavy odd proton nuclei belonging to the $\al$-decay
chain starting from $^{287}$115 has been investigated. RMF approach in the 
blocked BCS approximation gives very similar
structure for two forces NL1 and NL3. The quadrupole deformation increases 
smoothly from $^{287}$115 to $^{267}$Db.  The Q-values for probable decays
have been calculated and compared with experiment. The experimental values 
agree very well with the theoretical ones for the negative parity levels
calculated using the force NL3. Lifetime estimates obtained from 
phenomenological prescription indicate that the decays take place between
states having nearly identical quantum numbers.

\section*{Acknowledgements}
The calculations were done using the computer facilities provided by the
DSA Programme, University Grants Commission, New Delhi.

\clearpage
\begin{table}[ht]
\caption{Results of calculation in Md nuclei with using the forces NL1 and NL3.
The binding energy values are given in MeV. The quantities within the
parentheses are the calculated values of the energy levels in MeV. The 
experimental binding energy values are from Audi \etal\cite{AW}\label{be}}
\begin{tabular}{c|c|c|c|c|c}\hline
Nucleus&\multicolumn{1}{|c|}{Expt.} &\multicolumn{4}{|c}{Theo.}\\\cline{3-6}
&g.s.&\multicolumn{2}{|c}{NL1}&\multicolumn{2}{|c}{NL3}\\\cline{2-6}
&$J^\pi$&B.E.&\multicolumn{1}{|c|}{$J^\pi$(E$_{ex}$)}&
B.E.&\multicolumn{1}{|c}{$J^\pi$(E$_{ex}$)}\\
&B.E.&$\beta$&&$\beta_2$\\
\hline
$^{253}$Md&$\f{1}^-$&
1885.82&$\f{7}^-$(g.s.),$\f{1}^-$(0.02)&
1886.17&$\f{3}^-$(g.s.),$\f{7}^-$(0.18),\\
&1881.81$^*$&0.31&$\f{7}^+$(0.10),$\f{3}^-$(0.50)
&0.29&$\f{7}^+$(0.24),$\f{1}^-$(0.43)
\\\hline
$^{255}$Md&$\f{7}^-$&
1897.11&$\f{7}^-$(g.s.),$\f{1}^-$(0.01),&
1898.88&$\f{3}^-$(g.s.),$\f{7}^+$(0.27),\\
&1894.33&0.31&$\f{7}^+$(0.14),$\f{3}^-$(0.53)&
0.28&$\f{1}^-$(0.53),$\f{7}^-$(0.57) \\\hline
$^{257}$Md&$\f{7}^-$&
1907.68&$\f{7}^-$(g.s.),$\f{1}^-$(0.03),&
1910.42&$\f{3}^-$(g.s.),$\f{7}^+$(0.21),\\
&1906.32& 0.31&$\f{7}^+$(0.13) &
0.27&$\f{1}^-$(0.48),$\f{7}^-$(0.53)\\
\hline
\end{tabular}
$^*$Estimated value
\end{table}
\begin{table}
\caption{Calculated binding energy of the ground state of 
of nuclei belonging to the decay-chain of $^{287}$115.\label{be2}}
\begin{tabular}{ccc}\hline
Nucleus & \multicolumn{2}{c}{B.E.(MeV)}\\
&NL1&NL3\\\hline
$^{287}$115&2057.72&2055.84\\
$^{283}$113&2039.02&2038.87\\
$^{279}$111&2019.36&2020.65\\
$^{275}$Mt&2000.69&2002.03\\
$^{271}$Bh&1982.04&1983.22\\
$^{267}$Db&1962.15&1964.47\\\hline
\end{tabular}
\end{table}
\begin{table}
\caption{Calculated excitation energy  and quadrupole deformation of low-lying 
one quasiparticle states of nuclei belonging to the decay-chain of $^{287}$115. 
The states are denoted by Nilsson quantum nembers.\label{lev}}
\begin{tabular}{ccccccc}\hline
 &\multicolumn{3}{c}{NL1}&\multicolumn{3}{c}{NL3}\\
Nucleus & State & E(MeV) & $\beta_2$ & State & B(MeV) & $\beta_2$ \\\hline
$^{287}$115&
$\f{1}^-$[510]&0.00&0.165& 
$\f{1}^-$[510]&0.00&0.170\\
&$\f{7}^-$[503]&0.79&0.149&
$\f{3}^-$[512]&1.02&0.172\\
&$\f{3}^-$[512]&0.95&0.167&
$\f{11}^+$[615]&1.86&0.178\\
&$\f{11}^+$[615]&1.22&0.172&
$\f{9}^-$[505]&1.92&0.188\\\hline
$^{283}$113
&$\f{3}^-$[512]&0.00&0.182
&$\f{3}^-$[512]&0.00&0.178\\
&$\f{11}^+$[615]&0.21&0.187&
$\f{1}^-$[550]&0.72&0.167\\
&$\f{9}^-$[505]&0.58&0.200
&$\f{11}^+$[615]&0.76&0.182\\
&$\f{1}^-$[510]&0.65&0.187
&$\f{9}^-$[505]&1.26&0.191\\\hline
$^{279}$111
&$\f{3}^-$[512]&0.00&0.191
&$\f{11}^+$[615]&0.00&0.219\\
&$\f{9}^-$[505]&0.03&0.222
&$\f{5}^-$[512]&0.38&0.189\\
&$\f{11}^+$[615]&0.08&0.219
&$\f{3}^-$[512]&0.47&0.190\\
&$\f{1}^-$[510]&0.40&0.234
&$\f{1}^-$[550]&0.78&0.186\\
\hline
$^{275}$Mt
&$\f{3}^-$[512]&0.00&0.255
&$\f{5}^-$[512]&0.00&0.210\\
&$\f{9}^-$[505]&0.41&0.243
&$\f{11}^+$[615]&0.38&0.223\\
&$\f{11}^+$[615]&0.52&0.250
&$\f{9}^-$[505]&0.53&0.223\\
&$\f{1}^-$[510]&0.84&0.256
&$\f{1}^-$[521]&0.58&0.198\\
&$\f{5}^-$[512]&1.03&0.249
&$\f{9}^+$[624]&1.11&0.208\\\hline
$^{271}$Bh
&$\f{5}^-$[512]&0.00&0.271
&$\f{5}^-$[512]&0.00&0.263\\
&$\f{9}^+$[624]&0.53&0.270
&$\f{9}^-$[505]&0.33&0.247\\
&$\f{3}^-$[512]&1.08&0.269
&$\f{9}^+$[624]&1.08&0.259\\
&$\f{9}^-$[505]&1.13&0.259
&$\f{11}^+$[615]&1.11&0.256\\\hline
$^{267}$Db
&$\f{9}^+$[624]&0.00&0.282
&$\f{9}^+$[624]&0.00&0.274\\
&$\f{5}^-$[512]&0.58&0.280
&$\f{1}^-$[521]&0.30&0.268\\
&$\f{1}^-$[521]&0.83&0.276
&$\f{5}^-$[512]&0.99&0.269\\
&$\f{7}^+$[633]&1.27&0.278
&$\f{9}^-$[505]&1.39&0.258\\\hline
\end{tabular}
\end{table}
\begin{table}
\caption{Calculated and experimental Q-values for $\al$-decay. The subscript
$g$ to the levels indicate that the concerned level is the predicted ground 
state.\label{Q}}
\begin{tabular}{cllc}\hline
Parent&\multicolumn{2}{c}{Transition(Q-value in MeV)}&Q-value in MeV\\
\cline{2-4}
Nuclus & \multicolumn{2}{c}{Theo.}&\multicolumn{1}{c}{Expt.\cite{115}}\\\cline{2-3}
&\multicolumn{1}{c}{NL1}&\multicolumn{1}{c}{NL3}&\\\hline
$^{287}$115
&$\f{1}^-_g\ra\f{1}^-$(8.95)
&$\f{1}^-_g\ra\f{1}^-$(10.61)
&10.74$\pm$0.09\\
&$\f{1}^-_g\ra\f{3}_g^-$(9.60)
&$\f{1}^-_g\ra\f{3}_g^-$(11.33)&\\
&$\f{11}^+\ra\f{11}^+$(10.61)
&$\f{11}^+\ra\f{11}^+$(12.43)\\\hline
$^{283}$113
&$\f{3}^-_g\ra\f{3}_g^-$(8.64)
&$\f{3}^-_g\ra\f{3}^-$(9.61)&10.26$\pm$0.09\\
&
&$\f{3}^-_g\ra\f{5}^-$(9.70)&\\
&$\f{11}^+\ra\f{11}^+$(8.77)
&$\f{11}^+\ra\f{11}_g^+$(10.84)\\\hline
$^{279}$111
&$\f{3}^-_g\ra\f{3}_g^-$(9.63)
&$\f{5}^-\ra\f{5}_g^-$(10.06)&10.52$\pm$0.16\\
&$\f{11}^+\ra\f{11}^+$(9.19)
&$\f{11}^+_g\ra\f{11}^+$(9.30)\\\hline
$^{275}$Mt
&$\f{3}^-_g\ra\f{3}^-$(8.57)
&$\f{5}^-_g\ra\f{5}_g^-$(9.49)&10.48$\pm$0.09\\
&$\f{3}^-_g\ra\f{5}_g^-$(9.65)&
$\f{11}^+\ra\f{9}^+$(8.79)\\
&$\f{11}^+\ra\f{9}^+$(9.64)
&$\f{11}^+\ra\f{11}^+$(8.76)\\\hline
$^{271}$Bh
&$\f{5}^-_g\ra\f{5}^-$(7.83)
&$\f{5}^-_g\ra\f{5}^-$(8.56)&\\
&$\f{9}^+\ra\f{9}_g^+$(8.94)
&$\f{9}^+\ra\f{9}_g^+$(10.63)\\\hline
\end{tabular}
\end{table}
\begin{table}
\caption{Calculated and experimental lifetime for $\al$-decay. The theoretical
values have been iestimated using the $Q_\al$ values from the NL3 calculation
given in the first row of table \ref{Q} for each nucleus. See text for details.
\label{T}}
\begin{tabular}{ccc}\hline
Parent & $T_{\al th}$& $T_{\al ex}$\cite{115}\\\hline
\\
$^{287}$115 & 76.7 ms& 32$^{+155}_{-14}$ ms  \\
$^{283}$113 & 11.5 s & 100$^{+490}_{-45}$ ms\\
$^{279}$111 & 128.9 ms& 170$^{+810}_{-80}$ ms \\
$^{275}$Mt & 1.2 s & 9.7$^{+46}_{-4.4}$ s\\
$^{271}$Bh & 204.9 s& \\
\hline
\end{tabular}
\end{table}

\begin{thebibliography}{99}

\bibitem{115} Oganessian  Yu Ts \etal 2004 \PR C {\bf 69} 021601(R) 
\bibitem{H} Hoffmann U and Ring P 1988 \PL B {\bf  214} 307 
\bibitem{A} Afanasjev A,  K\"{o}nig J and Ring P 1996 \NP {\bf A608}
107 
\bibitem{Schuck} Ring P and Schuck P 1980 {\em The Nuclear
Many-Body Problem}, Springer -Verlag. 
\bibitem{ren}Ren Z, Chen D H, Tai F, Zhang H Y and  Shen W Q 2003
\PR C{\bf 67} 064302 
\bibitem{TMA} Ren Z and Toki H 2001  \NP {\bf A689} 691 
\bibitem{NLZ2} Bender M 2000 \PR C {\bf 61} 031302(R)
\bibitem{rash} Rashdan M 2002 \PR C{\bf 63} 064306 
\bibitem{geng} Geng L S, Toki H and  Meng J 2003 \PR C{\bf 68} 061303 
\bibitem{NL1}  Reinhard P -G,  Rufa M, Maruhn J A,  Greiner W and
Friedrich J 1986 \ZP A {\bf 323} 13 
\bibitem{NL3}  Lalazissis G A, K\"{o}nig J and Ring P 1997 \PR C{\bf 55}
540 
\bibitem{gamb} Gambhir Y K, Ring P and  Thimet A 1990 \APNY {\bf 198} 132 
\bibitem{ring} Ring P, Gambhir Y K and Lalazissis G A 1997
{\em Comput. Phys. Commun.}
{\bf 105} 77 
\bibitem{parna} Mitra P and Gangopadhyay G 2003 \PR C {\bf 68}
044319 
\bibitem{AW} Audi G, Wapstra A H and Thibault C 2003 \NP {\bf A729}  337
\bibitem{sh1} Patra S K, Greiner and Gupta R K 2000 \jpg {\bf 26} L65
\bibitem{sh2}  Reinhard P -G, Bender M and Maruhn J A 2002 {\em Comments in 
Mod. Phys.} {\bf 2} A117
\bibitem{VS} Viola V E and Seaborg G T 1966 {\em J. Inorg. Nucl. Chem.} 
{\bf 28} 741
\bibitem{M} Moeller P, Nix J R and Kratz K L 1997 {\em At. Data Nucl. Data Tables} 
{\bf 66} 131
\end{thebibliography}
\end{document}